# Scalable low loss cryogenic packaging of quantum memories in CMOS-foundry processed photonic chips

ROBERT BERNSON,[1,†,7] ALEX WITTE,[2,†,*] GENEVIEVE CLARK,[2,3] KAMIL GRADKOWSKI,[1] JEFFREY YANG,[2] MATT SAHA,[2] MATTHEW ZIMMERMANN,[2] ANDREW LEENHEER,[4] KEVIN C. CHEN,[3] GERALD GILBERT,[5] MATT EICHENFIELD,[4,6] DIRK ENGLUND,[3] AND PETER O'BRIEN[1]

[1]*Tyndall National Institute, Lee Maltings Complex, Dyke Parade, T12 R5CP, Cork, Ireland*
[2]*The MITRE Corporation, 202 Burlington Road, Bedford, MA 01730, USA*
[3]*Research Laboratory of Electronics, Massachusetts Institute of Technology, 50 Vassar Street, Cambridge, MA, 02139 USA*
[4]*Sandia National Laboratories, P.O. box 5800, Albuquerque, NM 87185, USA*
[5]*The MITRE Corporation, 200 Forrestal Road, Princeton, NJ 08540 USA*
[6] *College of Optical Science, University of Arizona, Tucson, AZ 85719 USA*
[†]*These authors contributed equally*
[7] *robert.bernson@tyndall.ie*
[*]*awitte@mitre.org*

**Abstract:** Optically linked solid-state quantum memories such as color centers in diamond are a promising platform for distributed quantum information processing and networking. Photonic integrated circuits (PICs) have emerged as a crucial enabling technology for these systems, integrating quantum memories with efficient electrical and optical interfaces in a compact and scalable platform. Packaging these hybrid chips into deployable modules while maintaining low optical loss and resiliency to temperature cycling is a central challenge to their practical use. We demonstrate a packaging method for PICs using surface grating couplers and angle-polished fiber arrays that is robust to temperature cycling, offers scalable channel count, applies to a wide variety of PIC platforms and wavelengths, and offers pathways to automated high-throughput packaging. Using this method, we show optically and electrically packaged quantum memory modules integrating all required qubit controls on chip, operating at millikelvin temperatures with < 3dB losses achievable from fiber to quantum memory.

## 1. Introduction

Recent progress in integrated optics has seen the use of optical fibers for scalable, low-loss packaging of PICs[1,2]. Quantum systems bring additional requirements for optical packaging, particularly approaches integrating solid-state single-photon sources and memories. Such systems typically require cryogenic temperatures (1), broad wavelength operation (2), and minimal photon loss (3). Optical packaging for these systems must also incorporate electrical co-packaging for qubit control and photon routing (4) while often accommodating high channel count (5) and heterogeneous integration of diverse materials (6)[3]. However, achieving broadband, efficient coupling between optical fibers and PIC waveguides involves significant engineering and precise alignment due to spatial mismatch between the respective optical modes[4]. Differences in thermal expansion among the optical fiber, PIC, and submounts can cause this alignment to shift and fail as devices are cooled to cryogenic temperatures, posing a major obstacle to achieving criteria 1-6.

These challenges have prompted exploration of diverse approaches to achieve effective optical integration of quantum PICs. Several demonstrations have shown cryogenic optical packaging of telecom wavelength PICs using single fibers coupled to grating couplers[5,6], but these approaches cannot accommodate scalable channel count. Multi-channel optically packaged PICs have recently been demonstrated at cryogenic temperatures using fiber arrays coupled to gratings; however, this relied on strict fabrication tolerances, strain relief structures, and vertical clearance due to the fiber array's angle of incidence[7]. Edge coupling to eight optical channels has been shown in visible wavelength PICs[8], and though this approach is promising, the requirements for mode matching between the PIC and optical fiber reduce the achievable coupling efficiency to heterogeneously integrated quantum memories. Additionally, this method requires painstaking manual alignment to achieve 3 dB coupling efficiency between fiber array and PIC. Less established coupling methods including tapered fibers[9–11] and photonic wire bonds[12] have recently led to improved coupling loss and broadband operation at visible wavelengths and cryogenic temperatures. These methods require custom equipment or assembly fiber-by-fiber however, hindering their practical scalability and prospects of eventual automation.

Here, we demonstrate a straightforward optical packaging approach using grating couplers and angle polished fiber arrays compatible with cryogenic operation and wavelengths ranging from visible to telecom. Our method uses off-the-shelf components in a process with an alignment tolerance of $\pm 2\mu m$ that is robust to repeated temperature cycling. This approach places no requirements on the photonic platform itself, allowing us to test over a dozen successful packages from two different foundry processes. We demonstrate operation of packaged visible and telecom grating coupled loopbacks at 6 K and room temperature with grating transmission losses per coupler as low as 2.2 dB for visible wavelengths, with potential for further improvement using advanced grating designs. In order to understand and design around the effects of thermal cycling, we also develop a comprehensive model of our fiber to grating interface which takes into account thermal, mechanical, and optical considerations. Finally, we apply our approach to foundry-process PICs optimized for low loss integration of diamond quantum memories. We show successful operation of our packaged PICs down to millikelvin temperatures with diamond to PIC coupling loss less than 0.5 dB. We demonstrate a fully electrically- and optically-packaged active quantum memory module in an active photonic chip with all required functionality for qubit control including on-chip optical routing and strain tuning.

## 2. Optical Packaging Method

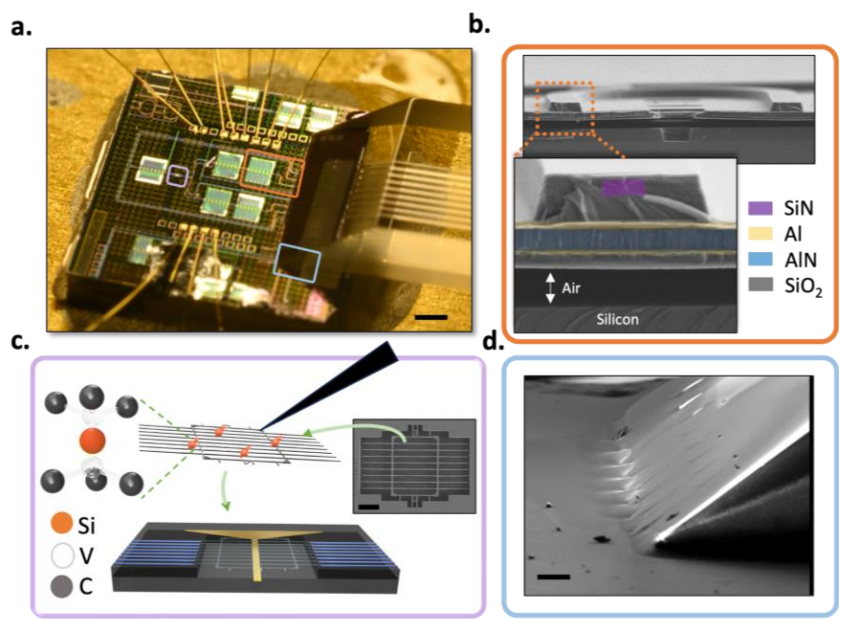

**Figure 1. a)** An optically and electrically packaged quantum PIC mounted on a copper submount and PCB with significant features noted in colored boxes and described in the rest of the figure. Scale bar is 500 μm. **b)** Cross-section scanning electron microscope (SEM) images of the piezoelectric modulators used in our photonic chips. **c)** Heterogeneous integration of color centers in diamond microchiplets with matching sockets in the PIC. **d)** SEM image of an angle-polished fiber array adhered to the PIC after optical packaging. Scale bar is 127 micrometers.

Figure 1 shows an overview of our packaging approach for an active, piezoelectric PIC with integrated quantum memories. The PIC[13,14] (Fig. 1a) is mounted on a copper submount for thermal contact with the cold finger of a cryostat and wirebonded to a printed circuit board (PCB) for electrical signal delivery to on-chip modulators and control lines. The piezoelectric platform itself (Fig. 1b) consists of silicon dioxide cladded silicon nitride waveguides, with aluminum layers delivering electrical signals to an aluminum nitride piezoelectric layer beneath the SiN waveguides. This platform allows both room temperature and cryogenic operation of stable, high extinction modulators for delivery and collection of photons in quantum PICs. We use a pick-and-stamp method[15] to detach diamond quantum microchiplets (QMCs) hosting group-IV color centers from a parent chip[16] and integrate them into sockets in the piezo-PIC (Fig. 1c). These sockets are equipped with exposed SiN waveguides for efficient optical coupling with the QMC waveguides[16,17], electrical lines for microwave spin control[18], and cantilevers for color center strain tuning[19]. Waveguide loopback structures are used to align an angle-polished fiber array with grating couplers on the PIC for optical excitation and fluorescence readout (Fig. 1d). Finally, we use UV-cured optical epoxy to attach the fiber array to the PIC, yielding a fully packaged, deployable quantum PIC (Fig. 1d).

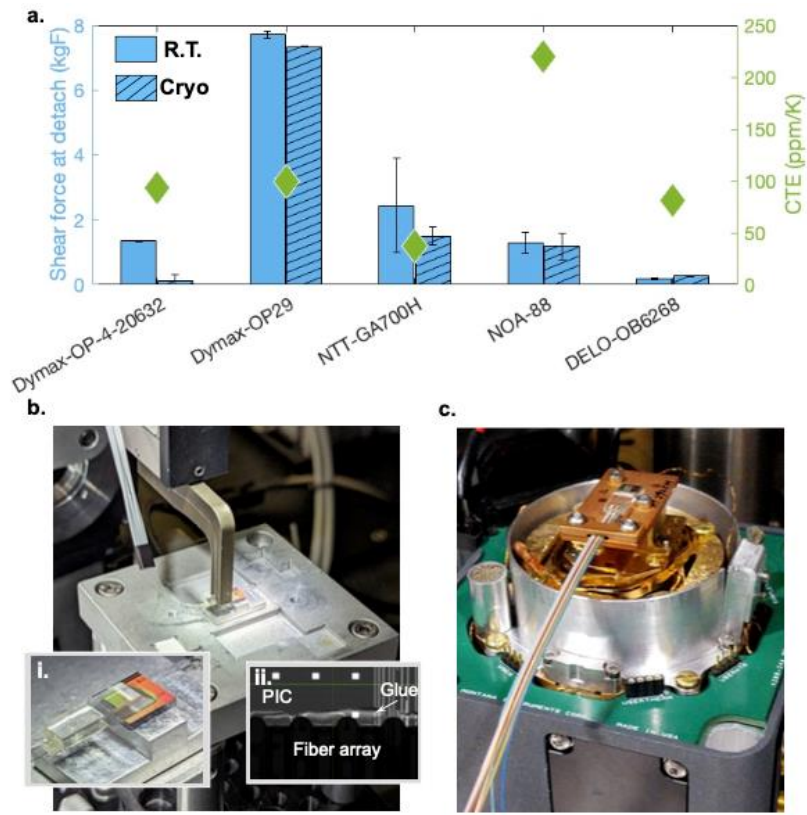

**Figure 2. a)** Average shear bond strength and CTE measurements for various commercially-available optical adhesives tested at room temperature and 7K. **b)** Reference package aligned and cured, i) fully packaged device released from grippers, ii) close-up surface image of fiber array attached to PIC via cured Dymax OP-29. **c)** A packaged module mounted on the cold finger of a Montana optical cryostat.

Adhesive selection plays a critical role in the success of our packaged modules at cryogenic temperatures. While commonplace in room-temperature photonic packaging, UV-cured optical adhesives are rarely tested at cryogenic temperatures required for quantum information processing[19]. As a first step in the development of our packaging process, the mechanical performance of five commercially available UV-cured adhesives were benchmarked in ambient conditions and at 7K. We selected Dymax OP-4-20632 due to its previously demonstrated use in room-temperature optical packaging[20], while the other four candidates were selected based on their advertised performance at 70K.

Since delamination and differences in the coefficient of thermal expansion (CTE) within the package can cause changes in optical alignment as a function of temperature, the chosen figures of merit for each epoxy are shear bond strength and CTE. Details of our testing regimen and tabulated results for shear bond strength are shown in Supplementary Section 1. These results, as well as CTE values at room temperature for each of the adhesives, are summarized in Figure 2a. The shear force required to detach the Dymax OP-29 sample is greater than 7 kgF with virtually no degradation after temperature cycling, nearly four times higher than the second best performing adhesive (DELO). The refractive index of cured Dymax

OP-29 is close to that of $SiO_2$ (1.50 vs ~1.46), and when compared to alternative optically transparent adhesives, it has a relatively short curing time[21]. Due to these properties, we selected Dymax OP-29 for fiber array attachment in our process.

After 6-axis active alignment between the fiber array and the PIC (Fig. 2b), a drop of OP-29 adhesive is placed between them, filling the air gap via capillary action [20]. This process confines a majority of the adhesive under the fiber array block, enabling close proximity to sensitive structures on the PIC surface like wirebonds and released MEMs devices (Fig. 2bii). After final adjustments to the fiber array's position, the output power from the device is continually monitored while the package surface is exposed to low-power UV light until the epoxy is fully cured and ready for cryogenic testing (Fig. 2c). Throughout our packaging process, we use automatic stages (Thorlabs NanoTrak) for active alignment optimization as well as an electronically-controlled syringe to dispense the epoxy, minimizing the need for human skill and intervention. Our process is robust enough for eventual full automation, making it attractive for circumstances where automated and high-throughput operation is desired.

3. **Results**

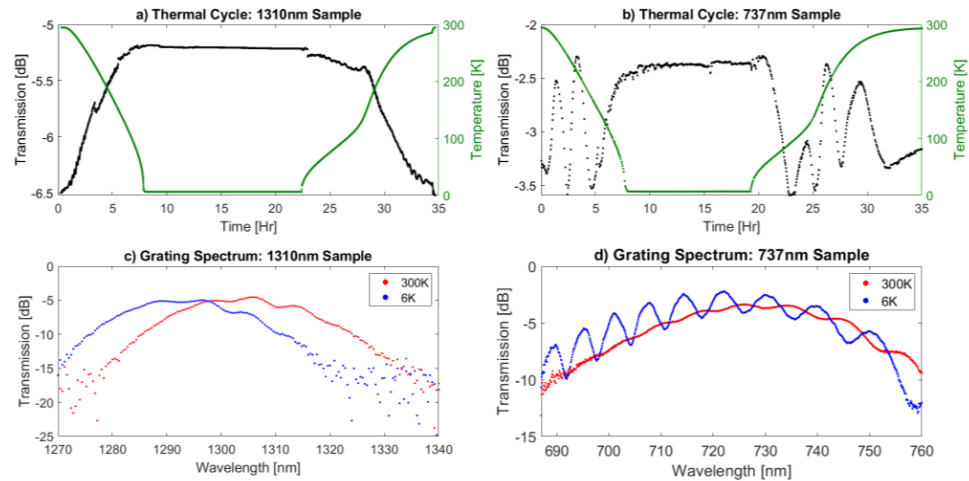

**Figure 3: Packaging performance under cryogenic conditions. a)** Thermal cycling of optically packaged 1310nm grating couplers, measured at 1300nm. After an angle polished fiber is glued to a grating loopback, the PIC is placed in a cryostat and cooled to 6K. Grating efficiency is measured during a full thermal cycle, holding 6K for a total of 15 hours. **b)** Thermal cycling of optically packaged 737nm grating couplers, measured at 730nm. During cycling, fiber shrinking causes polarization rotation, requiring constant re-optimization. **c)** Spectral response of the 1310nm gratings at room temperature as well as 6K. A 9nm blueshift is observed. **d)** Spectral response of the 737nm gratings at room temperature as well as 6K. A 1nm blueshift is observed.

Figure 3 shows results from two samples demonstrating our method's compatibility with cryogenic operation, one designed for 1310 nm operation (Fig. 3a,c) and the other for 730 nm (Fig. 3b,d). Both samples consist of PICs with grating coupled loopback waveguides and an angle polished fiber array glued to the chip. After fiber attachment, the packages are cooled to 6K while monitoring transmission at the output grating normalized to input power (Fig. 3a,b).

We observe slow polarization drift during cooldown which we ascribe to changes in the optical fiber as a function of temperature, and implement an automated polarization optimization setup (Supplementary Section 2) to counter this drift. After polarization correction, both the visible and telecom packages show improved transmission at the measurement wavelength at low temperatures compared to ambient conditions, with no failure or significant drop in transmission as the samples are cooled (Fig. 3a, b). We record optical spectra for each sample at room temperature and 6K and observe a reversible blueshift present in both samples upon cooling (Fig. 3c,d), as well as oscillations in the transmission of the visible package as a function of wavelength (Fig. 3d). We attribute these oscillations to Fabry-Perot resonances resulting from partial reflections between the bottom of the fiber block and top surface of the PIC, discussed in more detail below and in Supplementary Section 3. Both example packages undergo multiple thermal cycles in order to record these power and spectral measurements, with no indication of delamination or weakening in the adhesive provided the warming cycle does not shock the package with abrupt heater activation.

To understand the effects of temperature on our optical packaging, we construct a comprehensive model of the fiber array, adhesive, and visible grating couplers using finite element methods (COMSOL) (Fig. 4, inset). We first build the model in a 3D CAD environment to exactly match our package, combining the grating couplers and fiber array with a layer of adhesive between them such that the pitch of the fiber array matches our physical device at room temperature. We then use a thermal simulation to determine the geometry of our package at room temperature (300K) and cryogenic temperature (6K) to track any changes as the assembly is cooled. Finally, we use an optical simulation for the resulting geometry to determine the transmission spectrum for the package at room temperature (300K) and cryogenic temperature (6K) (Fig. 4b).

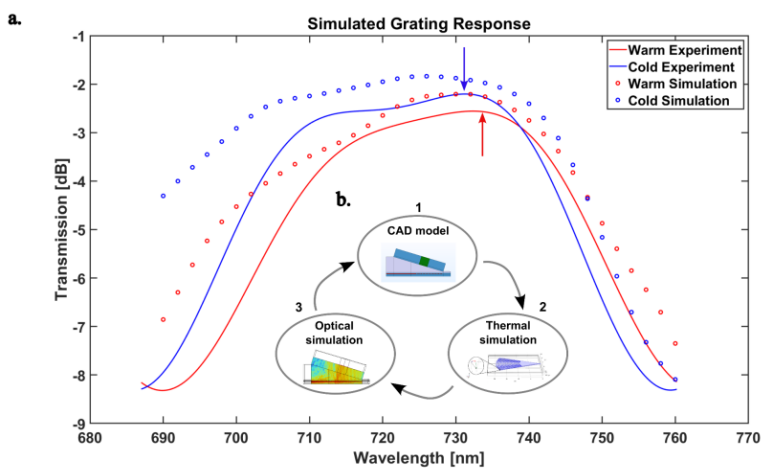

**Figure 4: a)** Simulated grating coupler spectral response at room temperature and 6 K. Oscillations due to Fabry-Perot effects have been removed. The experimentally measured 1nm shift in peak transmission wavelength at 6K corresponds to a 2° decrease in fiber pitch angle, as determined by optical simulations. **b)** Illustration of the simulation cycle used in this work. First, a model of the package is constructed. Thermal effects are then simulated and followed by optical spectrum simulations.

Figure 4a shows an overlay of the simulated (open circles) and measured (filled circles) responses at 6K and room temperature (we fit and remove the effects of Fabry-Perot resonances), indicating close agreement between the simulated and measured frequency shift in transmission. We measure a difference of 0.3 dB between the simulated maximum grating transmission and the experimentally measured fully packaged sample, indicating close alignment between simulation and our fabricated results. We observe through the thermal simulation that, due to the large CTE of the adhesive compared to the fiber and PIC, thicker sections of adhesive contract more than thinner ones during cooling. This results in a change in pitch of the fiber array if there is any initial variation in thickness of the adhesive layer. This change in pitch results in a temperature dependent change in incident angle of light onto the grating coupler, causing a corresponding shift in the transmission spectrum with temperature. We simulated an initial offset of 3° in the pitch angle of the fiber array corresponding to a +2° shift in the incident angle after cooldown. This shift in incident angle corresponds to a blueshift in the simulated optical spectra, consistent with our measurements. Since the shift in optimal wavelength seems to arise due to the slight change in incidence angle with temperature, we can improve our method to include either a pre-compensation shift in the incident angle prior to curing the adhesive, or variations in the grating structure to function optimally under the low temperature angle of incidence.

We also observe a change in period of the Fabry-Perot oscillations in the visible sample as it cools to 6K (Fig. 3d) which we attribute to the shortening distance between fiber array and PIC surfaces as the adhesive shrinks. By analyzing these oscillations as a Fabry-Perot cavity (Supplementary Section 3), we determine that the cavity distance at room temperature is $8.967 \mp 0.004 \mu m$ and $8.579 \mp 0.088 \mu m$ at 6K. These cavity lengths are consistent with our thermal simulations which estimate a 410 nm shrinkage of the adhesive layer between 300K and 6K.

## 4. Application to integrated quantum memories

We next apply our optical packaging approach to quantum memory modules consisting of diamond quantum microchiplets (QMCs) heterogeneously integrated into CMOS-process photonic chips. We use grating-coupled optical loopback structures (Fig. 5a) to establish the process and quantify optical losses arising from heterogeneous integration. QMCs hosting silicon vacancy color centers are integrated into sockets in the center of each loopback, where a window etch removes the oxide overcladding allowing direct contact between the QMC and SiN waveguides (Fig. 5ai, ii). We determine a loss in transmission as low as -0.03 dB at each window transition and -0.43 dB at each diamond-SiN transition for well aligned QMCs, based on transmission of a 737 nm laser (Supp. Fig. 8). With modest grating coupler performance (5 dB loss at 737 nm) this yields a fiber-PIC-QMC interface with less than 6 dB total loss, with the potential for losses under 3 dB with high performing grating couplers (see Fig. 3).

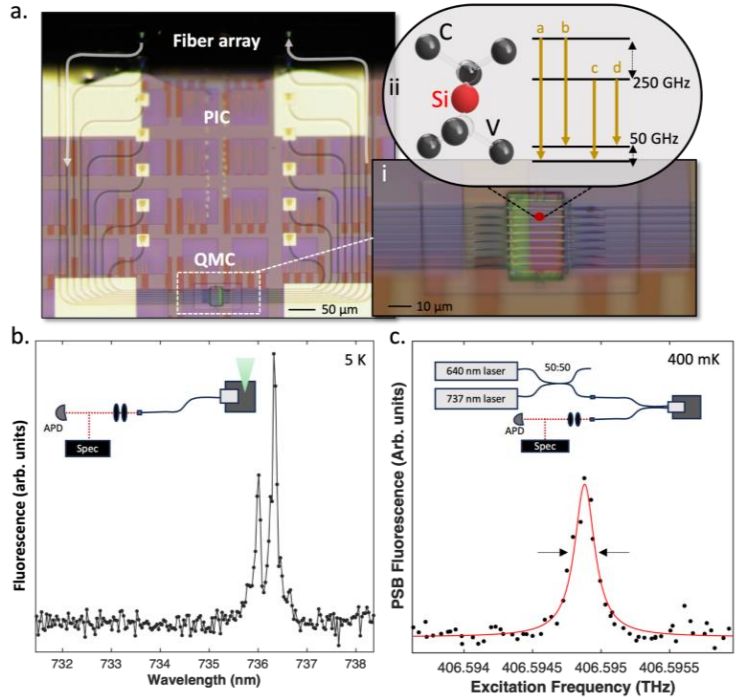

**Figure 5: Optically packaged diamond loopback chip measured at 400 mK. a)** Optical microscope image of a PIC with loopback structures connected through a heterogeneously integrated QMC. An angle polished fiber array glued to the PIC couples light into the PIC and collects fluorescence from silicon vacancy color centers in the QMC. **i)** Zoom-in of the QMC-PIC coupling region. Adiabatic tapering on the SiN and diamond provides high efficiency coupling between the SiN waveguides and aligned diamond nanobeams. **ii)** SiV structure, showing four spin-orbit split optical transitions. **b)** Photoluminescence spectrum measured from SiVs in the integrated QMC showing the SiV ZPL at 737 nm. **c)** Phonon sideband fluorescence from resonantly excited SiVs in the PIC, measured at 400 mK with on-chip excitation and collection.

After determining transmission efficiency, we measure fluorescence from silicon vacancy color centers (SiVs) in our optically packaged module under off-resonant excitation at 532 nm delivered in free-space, with the sample held at 5K in a Montana cryostat. We observe a clear signal from the SiV ZPL at 737 nm[21] collected through the on-chip waveguides and attached optical fiber (Fig. 5b). We next mount the same sample in a 400 mK sorption fridge with no free-space optical access and collect the phonon sideband (PSB) emission from SiVs excited resonantly through the on-chip waveguides. A fiber beam splitter combines a 640 nm laser with the resonant signal to provide alternating resonant and repump pulses. We observe multiple SiVs with near transform limited linewidths (Fig. 5c) within a single waveguide in the QMC.

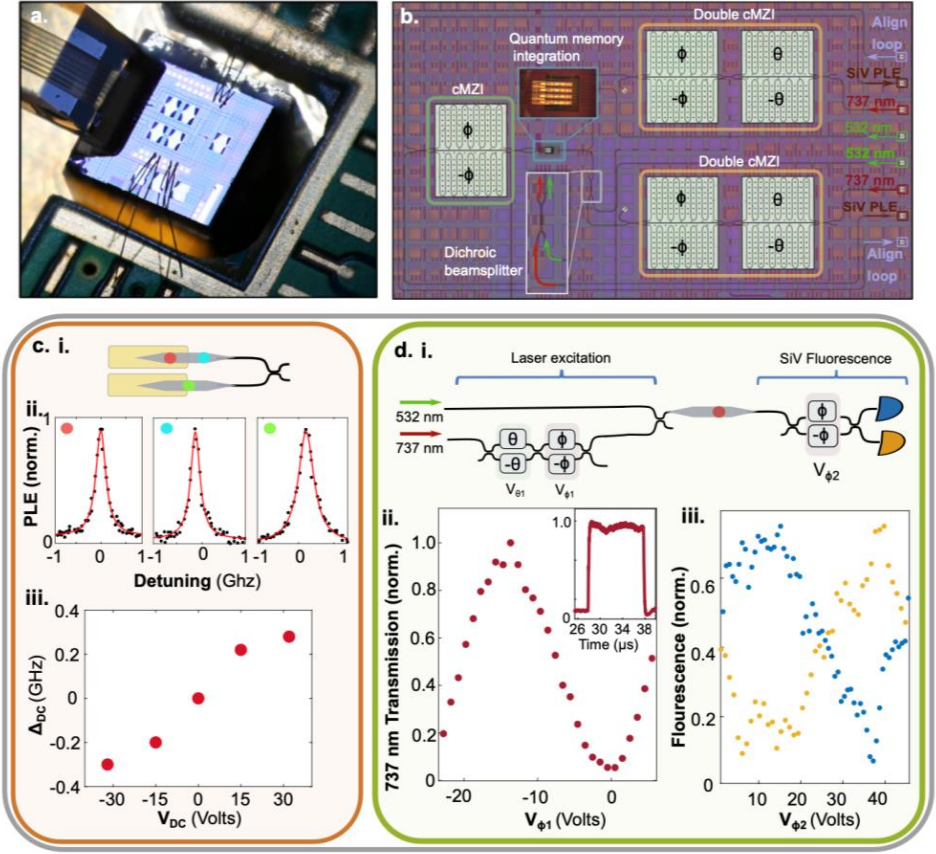

**Figure 6:** Electrically and optically packaged chip with integrated diamond. **a)** Optical image of the electrically and optically packaged PIC with integrated quantum memories. **b)** Zoom-in showing the QMC-integrated PIC, including cantilevers for strain control of quantum memories and optical modulators for control of excitation and collection optical paths. **c) i)** Sketch showing location of three SiVs in two channels of the integrated QMC. **ii)** PSB fluorescence from SiVs in channels 3 and 4 of the QMC collected through the attached fiber array. PLE measurements reveal high optical quality. **iii)** The integrated cantilevers allow frequency shifting of the ZPL transition to overcome inhomogeneities between color center transition energies. **d) i)** Sketch of device layout for a single channel in the QMC. Two input channels allow both off-resonant excitation or optical repump using a 532nm laser, and resonant excitation using 737 nm light. **ii)** A double cantilever MZI (cMZI) in the 737 nm path modulates the excitation power or generates optical pulses for resonant excitation. The two excitation channels are combined at a dichroic beamsplitter and directed into the QMC. **iii)** A cMZI at the output modulates fluorescence from SiVs between two detection channels for autocorrelation measurements and single photon routing.

Finally, we apply our optical packaging technique to an active, memory-integrated PIC with electrical packaging for color center control and on-chip optical modulation. Fig. 6a shows an image of the packaged PIC on a copper mount, wirebonded to a PCB for electrical signal routing. Within the PIC (Fig. 6b), double piezoelectric cantilever modulators[14] allow high contrast modulation of resonant laser input at 737 nm. A parallel input channel combines 532 nm laser input with the 737 nm path at a dichroic beamsplitter, directing both wavelengths into the QMC. SiV fluorescence is collected from the QMC into the on chip SiN waveguides and directed into a single cantilever modulator and onto two single photon detectors. Within the socket for the QMC, cantilevers mechanically contacting the diamond chiplet apply controllable strain to SiVs in the QMC for frequency tuning[19,22,23].

Figure 6c shows photoluminescence excitation (PLE) spectra from three SiVs in two channels of the QMC (i,ii) with frequency control to overcome inhomogeneity in their transition energies (Fig. 6cii). On-chip modulators and dichroic beamsplitters in the excitation path deliver resonant and off-resonant laser light to the SiVs without requiring a free-space beam or bulky off-chip modulators (Fig. 6di). This allows both off-resonant excitation at 532 nm (Supp. Fig. 10) and high-contrast modulation and pulsing of resonant 737 nm excitation (Fig. 6d ii). Applying a DC voltage to the cMZI in the collection path allows high contrast optical routing of SiV fluorescence between the two output ports (Fig. 6diii) and can function as an on-chip beamsplitter for coincidence counting or interference of photons from adjacent channels in the QMC. Our fiber packaging method allows integration of all optical modulation and qubit control components on chip without the need for active fiber alignment or free space optical access, while multiplexed excitation and readout channels allow scaling to large numbers of spin qubits in future work within a deployable, packaged module.

5. Conclusion

In conclusion, we demonstrate a robust and straightforward optical packaging approach that is compatible with cryogenic operation, heterogeneous integration, and electrical co-packaging. Grating couplers at visible and telecom wavelength allow for scalable optical packaging to arrays of single-mode fibers, while high shear-strength epoxy (Dymax OP-29) provides a fiber array-PIC bond that is resilient to repeated temperature cycling. Our approach does not require a thick oxide cladding for optimal mode matching which allows for the low-loss heterogeneous integration of quantum emitters and memories as well as other photonic materials[24–26]. This approach is also compatible with any foundry process that allows the construction of top-surface grating couplers and offers a pathway to automated, high throughput optical packaging of PICs. We demonstrate the utility of our approach by packaging CMOS foundry process PICs with heterogeneously integrated quantum memories hosted in diamond QMCs. Loopback structures reveal a fiber-PIC-memory interface with achievable losses below 3 dB while remaining compatible with electrical co-packaging and active photonic structures for routing and color center control. We demonstrate a fully packaged module with on chip resonant and off-resonant excitation, active routing of color center fluorescence, and strain tuning.

**Acknowledgements:**

This research was funded by Taighde Éireann - Research Ireland (formerly Science Foundation Ireland, SFI-12/RC/2276_P2).

Major funding for this work is provided by MITRE for the Quantum Moonshot Program. D.E. acknowledges partial support from Brookhaven National Laboratory, which is supported by the U.S. Department of Energy, Office of Basic Energy Sciences, under Contract No. DE-SC0012704 and the NSF RAISE TAQS program. Support is also acknowledged from the U.S.

Department of Energy, Office of Science, National Quantum Information Science Research Centers, Quantum Systems Accelerator.

M.E. performed this work, in part, with funding from the Center for Integrated Nanotechnologies, an Office of Science User Facility operated for the U.S. Department of Energy Office of Science.

Sandia National Laboratories is a multimission laboratory managed and operated by National Technology & Engineering Solutions of Sandia, LLC, a wholly owned subsidiary of Honeywell International Inc., for the U.S. Department of Energy's National Nuclear Security Administration under contract DE-NA0003525. This paper describes objective technical results and analysis. Any subjective views or opinions that might be expressed in the paper do not necessarily represent the views of the U.S. Department of Energy or the United States Government.

**Competing interest:** D.E. is a Scientific Advisor to and holds shares in QuEra Computing. The other authors declare no competing interests.